\documentclass[twocolumn,superscriptaddress,longbibliography]{revtex4-2}
\usepackage{hyperref,amsmath,amssymb,tikz,scalerel}
\hypersetup{
        colorlinks=true,
        linkcolor=red,
        citecolor=blue,
        urlcolor=blue}
\usetikzlibrary{svg.path}
\definecolor{orcidlogocol}{HTML}{A6CE39}
\tikzset{
	orcidlogo/.pic={
		\fill[orcidlogocol] svg{M256,128c0,70.7-57.3,128-128,128C57.3,256,0,198.7,0,128C0,57.3,57.3,0,128,0C198.7,0,256,57.3,256,128z};
		\fill[white] svg{M86.3,186.2H70.9V79.1h15.4v48.4V186.2z}
		svg{M108.9,79.1h41.6c39.6,0,57,28.3,57,53.6c0,27.5-21.5,53.6-56.8,53.6h-41.8V79.1z M124.3,172.4h24.5c34.9,0,42.9-26.5,42.9-39.7c0-21.5-13.7-39.7-43.7-39.7h-23.7V172.4z}
		svg{M88.7,56.8c0,5.5-4.5,10.1-10.1,10.1c-5.6,0-10.1-4.6-10.1-10.1c0-5.6,4.5-10.1,10.1-10.1C84.2,46.7,88.7,51.3,88.7,56.8z};}}
\newcommand\orcid[1]{\href{https://orcid.org/#1}{\mbox{\scalerel*{\begin{tikzpicture}[yscale=-1,transform shape]\pic{orcidlogo};\end{tikzpicture}}{|}}}}

\begin{document}
\title{Strong coupling of quantum emitters and the exciton polariton in MoS$_2$ nanodisks}
\author{Feng-Zhou Ji}
\affiliation{Lanzhou Center for Theoretical Physics, Key Laboratory of Theoretical Physics of Gansu Province, Lanzhou University, Lanzhou 730000, China}
\author{Si-Yuan Bai\orcid{0000-0002-4768-6260}}
\affiliation{Lanzhou Center for Theoretical Physics, Key Laboratory of Theoretical Physics of Gansu Province, Lanzhou University, Lanzhou 730000, China}
\author{Jun-Hong An\orcid{0000-0002-3475-0729}}
\email{anjhong@lzu.edu.cn}
\affiliation{Lanzhou Center for Theoretical Physics, Key Laboratory of Theoretical Physics of Gansu Province, Lanzhou University, Lanzhou 730000, China}

\begin{abstract}
As a quasiparticle formed by light and excitons in semiconductors, the exciton-polariton (EP) as a quantum bus is promising for the development of quantum interconnect devices at room temperature. However, the significant damping of EPs in the material generally causes a loss of quantum information. We propose a mechanism to overcome the destructive effect of a damping EP on its mediated correlation dynamics of quantum emitters (QEs). Via an investigation of the near-field coupling between two QEs and the EP in a monolayer MoS$_{2}$ nanodisk, we find that, with the complete dissipation of the QEs efficiently avoided, a persistent quantum correlation between the QEs can be generated and stabilized even to their steady state. This is due to the fact that, with upon decreasing the QE-MoS$_2$ distance, the QEs become so hybridized with the EP that one or two bound states are formed between them. Our result supplies a useful way to avoid the destructive impact of EP damping, and it refreshes our understanding of the light-matter interaction in absorbing medium.
\end{abstract}
\maketitle

\section{\label{sec:level1}Introduction}
Two-dimensional (2D) materials have become an exciting platform for strong light-matter interaction enabled by their polaritonic modes \cite{Basov2016,nmat4792,Zhang2021,doi:10.1063/5.0074355}. As a mixture of electromagnetic field (EMF) and polarized elements, various polaritonic modes have been realized in different 2D materials, including the plasmon polariton in graphene \cite{Koppens2011,Grigorenko2012,Hu2022}, the phonon polariton in hexagonal boron nitride \cite{Caldwell2014,Caldwell2019,AlfaroMozaz2019}, and the exciton polariton (EP) in transition-metal dichalcogenide monolayers \cite{Dufferwiel2015,nphoton.2014.304,srep33134,Schneider2018,acsphotonics.0c00063,PhysRevB.103.235409}. The feature of the subwavelength and subdiffraction light confinement of different types of polaritons endows these systems with a desired near-field enhancement of light-matter interaction \cite{PhysRevLett.114.157401,s41467-021-26617-w}. It makes polaritons potential candidates for developing novel polaritonic devices with improved functionalities that span from sensing, nonlinear optics, and photovoltaics to quantum technology \cite{nmat3950,Sanvitto2016,Shi2017,s42247-021-00200-x,Yu2021}. In the polariton family, the EP exhibits good performance at room temperature due to the large exciton-binding energy \cite{srep33134}.

Strong light-matter coupling in 2D materials is significantly damped by the material absorption, which leads to short-lived polariton modes \cite{PhysRevB.97.205436,PhysRevLett.126.075301} and severely hinders their applications.
It was found that, when a quantum emitter (QE) is placed in proximity to a nanoscale material, its spontaneous emission rate is enhanced several orders of magnitude compared with its free-space values by the polaritonic modes \cite{PhysRevLett.112.253601,Toermae2014,PhysRevB.93.035426,Baranov2018,PhysRevB.100.245403,PhysRevB.102.075446,PhysRevResearch.2.033141,PhysRevB.103.035416}. This is known as the Purcell effect established under the Markovian approximation, which generally works in the weak coupling limit. Obviously, such a dramatically enhanced coupling between the QEs and polaritonic modes invalidates the physical condition under which the Markovian approximation is applicable \cite{PhysRevB.82.115334,PhysRevB.103.085407}. It was really found that a fast QE-polariton energy-exchange appears in the non-Markovian dynamics when their distance is small \cite{PhysRevB.89.041402,PhysRevB.95.075412,PhysRevB.99.195412,PhysRevB.104.L201405}. However, this reversible dynamics tends to vanish in the long-time limit with the QE asymptotically decaying to its ground state due to the damping nature of the polariton modes. How to suppress the destructive effect of the finite-lifetime EP on the QE is significant in the fundamental physics of quantum control. Furthermore, in practical quantum technology with the polariton as a quantum bus to distribute quantum coherence between the separated QEs \cite{PhysRevLett.106.020501,PhysRevLett.110.080502,PhysRevLett.115.163603,PhysRevResearch.1.023027,PhysRevLett.127.083602}, one always desires that the decoherence of the QEs caused by the damping EP could be suppressed. Therefore, how to control the influence of the damping of the EP on the QEs is a key problem.

\begin{figure}[tbp]
\includegraphics[width=1.0\columnwidth]{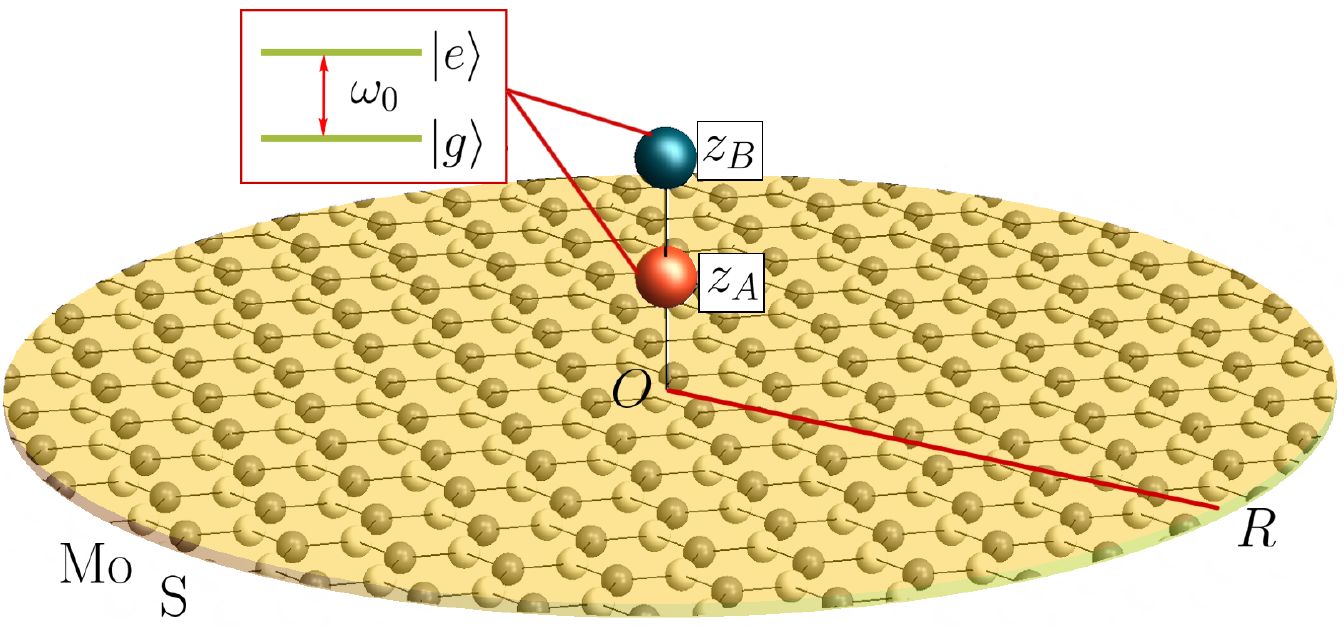}
\caption{ Schematic illustration of the system. Two QEs, with identical dipole moment ${\bf d}_{A/B}=d\hat{\bf e}_x$ and eigenfrequency $\omega_{A/B}=\omega_0$, are located at $\textbf{r}_{A/B}=(0,0,z_{A/B})$ along the central axis of a MoS$_{2}$ nanodisk with radius $R$. The radiative fields of the QEs induces an EP on the MoS$_{2}$ nanodisk, which in turn mediates an effective coupling between the QEs.    }\label{sketch}
\end{figure}
Here, going beyond the Markovian approximation, we perform a fundamental scientific study on the correlated dynamics of two separated QEs mediated by a common EP on a MoS$_{2}$ nanodisk. It is found that, quite different from the exclusively complete decay in previous works, the exact dynamics of the QEs exhibits diverse behaviors, including complete decay, population trapping, and persistent Rabi-like oscillation in different QE-nanodisk distances. Our analysis reveals that they are essentially determined by different numbers of bound states formed by the total system of the QEs and EP. Supplying a novel mechanism to suppress the destructive effect of the damping EP on the QEs by engineering the formation of the bound states, our result might be helpful to design quantum polaritonic devices in the MoS$_2$ nanostructure.

\section{System and quantization}
We investigate the strong quantized light-matter interactions between two QEs and the emitted EMF propagating in a monolayer MoS$_{2}$ nanodisk (see Fig. \ref{sketch}). The EMF triggers three distinct modes in our considered structure. The first one is the radiative mode propagating into the vacuum dielectric. The second one is the damped nonradiative mode absorbed by the MoS$_{2}$ nanodisk. The last one is the tightly confined hybrid mode of the EMF and the exciton in the MoS$_2$ called an EP propagating along the MoS$_2$-dielectric interface. The EP enables a tight confinement of light on the interface and thus enhances the strong coupling between the QEs and the EMF. The absorption nature of the 2D material to the EMF endows it with a substantial difference from the conventional light-matter interactions in a dispersion medium. A macroscopic quantization method of EMF in the absorbing medium has been proposed based on the dyadic Green's tensor, where the absorption of the medium to light is described by a Langevin noise \cite{PhysRevA.51.3246,PhysRevA.57.3931,PhysRevB.93.035426,PhysRevB.94.195418,PhysRevB.103.174421}. Then the electric field reads
\begin{equation}
\hat{\mathbf{E}}(\mathbf{r},\omega )=\frac{ic^{-2}\omega ^{2}}{\sqrt{\pi
\varepsilon _{0}/\hbar }}\int d^{3}\mathbf{r}'\sqrt{\text{Im}[\varepsilon(\omega )]}\mathbf{G}(\mathbf{r},\mathbf{r}',\omega )\cdot \mathbf{\hat{f}}(\mathbf{r}',\omega ),\nonumber
\end{equation}
where $\varepsilon_{0}$ is the vacuum permittivity, $\varepsilon(\omega)$ is the dielectric permittivity of MoS$_2$, $c$ is the speed of light in a free space, and $\hat{\bf{f}}(\bf{r},\omega)$ satisfying $[\hat{\bf{f}}(\bf{r},\omega),\hat{\bf{f}}^\dag(\bf{r}',\omega')]=\delta(\bf{r}-\bf{r}')\delta({\omega-\omega'})$ is the annihilation operator of the EP. The Green's tensor $\bf{G(\bf{r},\bf{r}^{\prime},\omega)}$ is rendered as the field in frequency $\omega$ evaluated at $\bf{r}$ due to a point source at $\bf{r}^{\prime}$, and it satisfies the Maxwell-Helmholtz equation $[{\pmb\nabla}\times{\pmb\nabla}\times-\omega^{2}\varepsilon(\omega)/c^{2}]\mathbf{G}(\mathbf{r},\mathbf{r}^{\prime },\omega )=\mathbf{I}\delta (\mathbf{r}-\mathbf{r}^{\prime })$, where $\mathbf{I}$ is the identity matrix. Incorporating the dispersion and the loss of the EMF in the MoS$_2$-dielectric interface into the Green's tensor, such a quantization scheme takes good care of the medium absorption and dispersion to the EP. It allows for a complete description of the quantized light-matter interaction by calculating $\bf{G(\bf{r},\bf{r}^{\prime},\omega)}$ appropriately for the corresponding nanostructures.

We consider that the two QEs located at the central axis of the MoS$_{2}$ nanodisk, i.e., $\textbf{r}_{l}=(0,0,z_{l})$ $(l=A,B)$, interact independently with the EP. Under the dipole and rotating-wave approximations, the Hamiltonian reads
\begin{eqnarray}\label{Hmt}
\hat{H} &=&\sum_{l=A,B}\hbar \omega _{l}\hat{\sigma}^{\dag}_{l}\hat{\sigma}_{l}+\int d^{3}\mathbf{r}\int d\omega \hbar \omega \hat{\mathbf{f}}^{\dag }(\mathbf{r},\omega )\cdot \hat{\mathbf{f}}(\mathbf{r},\omega )\nonumber \\
&&-\sum_{l=A,B}\int d\omega \lbrack \mathbf{\bf d }_{l}\cdot \hat{\mathbf{E}}(\mathbf{r}_{l},\omega )\hat{\sigma}^{\dag}_{l}+\text{H.c.}],
\end{eqnarray}
where $\hat{\sigma}_{l}=|g_{l}\rangle \langle e_{l}|$ and $\mathbf{d}_l$ are the transition operator and electric dipole moment between the excited state $|e_{l}\rangle$ and the ground state $|g_{l}\rangle$ with frequency $\omega_l$ of the $l$th QE. The validity of the rotating-wave approximation in this system has been investigated \cite{PhysRevB.97.115402}. The dipole approximation is valid when the size of the QEs, which is about 1 nm for the $J$ aggregates \cite{PhysRevLett.118.237401} and 20 nm for semiconductor quantum dots \cite{PhysRevLett.114.247401}, is sufficiently smaller \cite{PhysRevB.82.115334,PhysRevB.86.085304} than the wavelength of the EP, which is about 500 nm in the transition metal dichalcogenide \cite{Hu2017}. Similar models were studied in Refs. \cite{PhysRevB.94.165302,PhysRevLett.106.020501,PhysRevB.94.195418,PhysRevB.103.085407,PhysRevB.103.174421,PhysRevA.96.012339}, but they are under the Markovian approximation. The non-Markovian dynamics of one single QE interacting with different polariton modes was considered only in recent years \cite{PhysRevB.100.245403,PhysRevB.99.195412,PhysRevB.102.075446,PhysRevResearch.2.033141,doi:10.1063/5.0074355,PhysRevB.103.035416,PhysRevB.104.L201405,PhysRevA.106.013718}.

\section{Non-Markovian Dynamics}
One can find that the total excitation number operator $\hat{N}=\sum_{l}\hat{\sigma}^{\dag}_{l}\hat{\sigma}_{l}+\int d^{3}\mathbf{r}\int d\omega \mathbf{\hat{f}}^{\dag }(\mathbf{r},\omega )\cdot \mathbf{\hat{f}}(\mathbf{r},\omega )$ is conserved. Consider the situation in which the QEs reside in a separable state and the EP is in the vacuum state $|\{0_{{\bf r},\omega}\}\rangle $ initially, i.e., $|\Psi(0)\rangle=|e_A,g_B;\{0_{{\bf r},\omega}\}\rangle$. The evolved state is then expanded as $|\Psi(t)\rangle=\big[\sum_{l}c_{l}(t)\hat{\sigma}_l^\dag+\int d^{3}\mathbf{r}\int d\omega b_{{\bf r},\omega}(t)\hat{\mathbf{f}}^\dag(\mathbf{r},\omega)\big]|g_A,g_B;\{0_{{\bf r},\omega}\}\rangle$. It can be derived from the time-dependent Schr\"{o}dinger equation that the probability amplitudes $c_{l}(t)$ satisfy
\begin{equation}
\dot{c}_{l}(t)+i\omega _{l}c_{l}(t)+\sum_{j=A,B}\int_{0}^{t}d\tau f_{lj}(t-\tau )c_{j}(\tau )=0 \label{evolution}
\end{equation}
under $c_A(0)=1$ and $c_B(0)=0$. Here $f_{lj}(x)=\int_{0}^{\infty}d\omega J_{lj}(\omega )\exp(-i\omega x)$ are the correlation functions and $J_{lj }(\omega )=\omega ^{2}\mathbf{d }_{l}\cdot\textrm{Im}[ \mathbf{G}(\mathbf{r}_{l}, \mathbf{r}_{j},\omega )]\cdot \mathbf{d }_{j}^{\ast }/(\pi\hbar \varepsilon_{0}c^{2})$ are the spectral densities incorporating the actions of the EP on the QEs. In obtaining Eq. \eqref{evolution}, we have used the relation $\int d^3\mathbf{s}\frac{\omega ^{2}}{c^{2}}$Im$[\varepsilon (\omega )]\mathbf{G}(\mathbf{r},\mathbf{s},\omega )\mathbf{G}^{\ast }(\mathbf{r}^{\prime },\mathbf{s},\omega )=\textrm{Im}[\mathbf{G}(\mathbf{r},\mathbf{r}^{\prime },\omega )]$ \cite{PhysRevA.57.3931,PhysRevB.95.161408}. The convolution in Eqs. \eqref{evolution} reflecting the memory effect renders the dynamics of two QEs interacting with the common EP non-Markovian. The correlation between $c_A(t)$ and $c_B(t)$ in Eqs. (\ref{evolution}) indicates that although there is no direct interaction between the two QEs in Eq. (\ref{Hmt}), a coherent coupling between them is effectively induced by the EP. The Green's tensor $\mathbf{G}(\mathbf{r},\mathbf{r}^{\prime },\omega )$ for the 2D monolayer nanodisk of our system is analytically solvable, from which the spectral densities $J_{lj}(\omega )$ can be obtained. To facilitate the analysis, we consider that the QEs have identical eigenfrequency $\omega_A=\omega_B\equiv\omega_{0}$ and parallel dipole moment along the $x$ axis ${\bf d}_{A}={\bf d}_{B}\equiv d\hat{\mathbf{e}}_{x}$. Then only the $xx$-component of the Green's tensor contributes to the spectral densities $J_{lj}(\omega )=\frac{3\Gamma_{0}\omega^{2} c}{\omega_{0}^{3}} \text{Im}[G_{xx}(z_l,z_j,\omega)]$, with \cite{PhysRevB.93.035426}
\begin{equation}
G_{xx}(z,z',\omega)=\frac{-c^{2}}{2\omega^{2}}\sum_{n=0}^{\infty}\alpha_{n}^{1}(z^{\prime},
\omega)\frac{[g(z)-z/R]^{2n+2}}{g(z)},\label{spec}
\end{equation}
where $g(z)=\sqrt{(z/R)^{2}+1}$, $\Gamma_{0}=\omega_{0}^{3}d^{2}/(3\pi\hbar\varepsilon_{0}c^{3})$ is the spontaneous emission rate in the free space characterizing the intrinsic lifetime of the QEs, and the forms of $\alpha_{n}^{1}(z^{\prime},\omega)$ are given in Appendix \ref{appgf}. With Eq. \eqref{spec}, the dynamics of the QEs is obtained by numerically solving Eqs. \eqref{evolution}.

When the coupling between the QEs and the EP is weak and the time scale of the correlation functions of the EP is much smaller than that of the QEs, we can safely apply the Markovian approximation by neglecting the memory effect in Eqs. \eqref{evolution}. Then their Markovian approximate solutions read (see Appendix \ref{appmark})
\begin{eqnarray}
c^\text{MA}_{A}(t) &=&e^{-[i\omega _{0}+(\Upsilon _{AA}+\Upsilon _{BB})]t}[\frac{\Upsilon _{BB}-\Upsilon _{AA}}{\Omega }\sinh (\Omega t)\nonumber \\
&&+\cosh(\Omega t)],\label{cat}\\
c^\text{MA}_{B}(t) &=&-2e^{-[i\omega _{0}+(\Upsilon _{AA}+\Upsilon _{BB})]t}\frac{%
\Upsilon _{BA}}{\Omega }\sinh (\Omega t),\label{cbt}
\end{eqnarray}
where $\Upsilon _{ij}=\frac{\gamma _{ij}}{2}+i\frac{\Delta _{ij}}{2}$, $\Omega =[(\Upsilon _{AA}-\Upsilon _{BB})^{2}+4\Upsilon _{AB}\Upsilon_{BA}]^{1/2}$, $\gamma _{ij}=\pi J_{ij}(\omega _{0})$, and $\Delta_{ij}=\mathcal{P}\int d\omega {\frac{J_{ij}(\omega )}{\omega _{0}-\omega }}$. Here $\gamma_{ll}$ and $\gamma_{lm}$ are the individual and correlated spontaneous emission rates, and $\Delta_{ll}$ and $\Delta_{lm}$ are the Lamb shift and dipole-dipole interaction strength of the two QEs induced by the EP. It is found that the reversible energy exchange between the two QEs mediated by the EP tends to vanish in the long-time limit due to the exponential decay in the global factor of Eqs. \eqref{cat} and \eqref{cbt}. Being consistent with the previous works \cite{PhysRevB.94.165302,PhysRevA.96.012339,PhysRevLett.106.020501,PhysRevB.103.085407,PhysRevB.94.195418,PhysRevB.103.085407,PhysRevB.103.174421}, this manifests the destructive effect of the damping EP on the QEs.

Quite different from the approximate result, we will show that such mediated coherent coupling can induce a persistently reversible energy-exchange even in the steady state when the Markovian approximation is relaxed. The formal solutions of Eqs. \eqref{evolution} in the general non-Markovian dynamics are obtainable by using the Laplace transform method. It reduces Eqs. \eqref{evolution} to
\begin{eqnarray}
\tilde{c}_l(s)&=&\frac{\zeta_l(s)}{\Xi _{A}(s)\Xi _{B}(s)-\tilde{f}_{AB}(s)\tilde{f}_{BA}(s)},\label{cldtts}
\end{eqnarray}
where $\Xi _{l}(s)=s+i\omega _{0}+\tilde{f}_{ll}(s)$, $\tilde{f}_{lj}(s)=\int_{0}^{\infty } {J_{lj}(\omega)d\omega\over s+i\omega}$, $\zeta_A(s)=\Xi_{B}(s)$, and $\zeta_B(s)=-\tilde{f}_{BA}(s)$. Then $c_l(t)$ are obtained by the inverse Laplace transform to $\tilde{c}_{l}(s)$, which can be done by finding the poles from
\begin{equation}
\Xi _{A}({E\over i\hbar})\Xi _{B}({E\over i\hbar})-\tilde{f}_{AB}({E\over i\hbar})\tilde{f}_{BA}({E\over i\hbar})=0,\label{eigroot}
\end{equation}
where we have made the replacement $s=-iE/\hbar$ for convenience. It is interesting to find that the roots $E$ of Eq. \eqref{eigroot} are just the eigenenergies of Eq. \eqref{Hmt}. To prove this, we expand the eigenstate as $|\Phi\rangle=\big[\sum_{l}x_l\hat{\sigma}_l^\dag+\int d^{3}\mathbf{r}\int d\omega y_{{\bf r},\omega}\hat{\mathbf{f}}^\dag(\mathbf{r},\omega)\big]|g_A,g_B;\{0_{{\bf r},\omega}\}\rangle$. The substitution of it into $\hat{H}|\Phi\rangle=E|\Phi\rangle$ results in the coupled equations of $x_A$ and $x_B$ as
\begin{equation}
(E-\hbar\omega_0)x_l=-i\hbar\sum_{j=A,B}\tilde{f}_{lj}(-iE/\hbar)x_j,
\end{equation}which readily lead to Eq. \eqref{eigroot} after eliminating $x_l$ (see Appendix \ref{appenspe}). Therefore, on the one hand, Eq. \eqref{eigroot} governs the behavior of $c_l(t)$, while on the other hand, it determines the eigenenergies of the total system. This implies that the dynamics of the QEs is essentially determined by the energy-spectrum characteristic of the total system. The left hand side of Eq. \eqref{eigroot} is ill-defined and jumps rapidly between $\pm\infty$ in the regime $E>0$ because of the divergence of the integrand in $\tilde{f}_{lj}(-iE/\hbar)$. Thus Eq. \eqref{eigroot} has infinite positive roots, which form a continuous energy band. However, in the regime $E<0$, the left hand side of Eq. \eqref{eigroot} is a quadratic-like function of $E$. Therefore, depending on the system parameters, at most two isolated roots $E_b^j$ of Eq. \eqref{eigroot} exist in this regime. We call the eigenstates corresponding to these isolated eigenenergies $E_b^j$ bound states. Substituting these poles into the inverse Laplace tansform, we find that they contribute two nontrivial residues, which are preserved in the long-time limit; while the continuous energy band contributes a branch cut, which tends to zero in the long-time limit due to the out-of-phase interference. We thus have
\begin{equation}
\lim_{t\rightarrow\infty}c_l(t)=\begin{cases}0, ~~~~~~~~~~~~~~~~~~~~\text{no bound state},\\ \sum_{j=1}^MZ_l^je^{{-i\over\hbar}E_b^jt}, ~M~\text{bound states},\end{cases}\label{ltmcl}
\end{equation}
where $M$ is the number of the formed bound states, and $Z_l^j=\frac{\zeta _{l}(s)}{\partial_s[\Xi _{A}(s)\Xi _{B}(s)-
\tilde{f}_{AB}(s)\tilde{f}_{BA}(s)]}\big|_{s=-iE_{b}^{j}/\hbar }$ is the residue of the $j$th bound state $E_b^j$ contributing to $c_l(t)$ (see Appendix \ref{appsssn}). Equation \eqref{ltmcl} indicates the decisive role played by the energy-spectrum characteristic of the total QE-EP system in the non-Markovian dynamics of the QEs. It is remarkable to see that the formation of the bound states prevents $|c_l(t)|^2$ from decaying to zero. This result is not captured by the Markovian approximation in Eqs. \eqref{cat} and \eqref{cbt}.

\begin{figure}[tbp]
\includegraphics[width=\columnwidth]{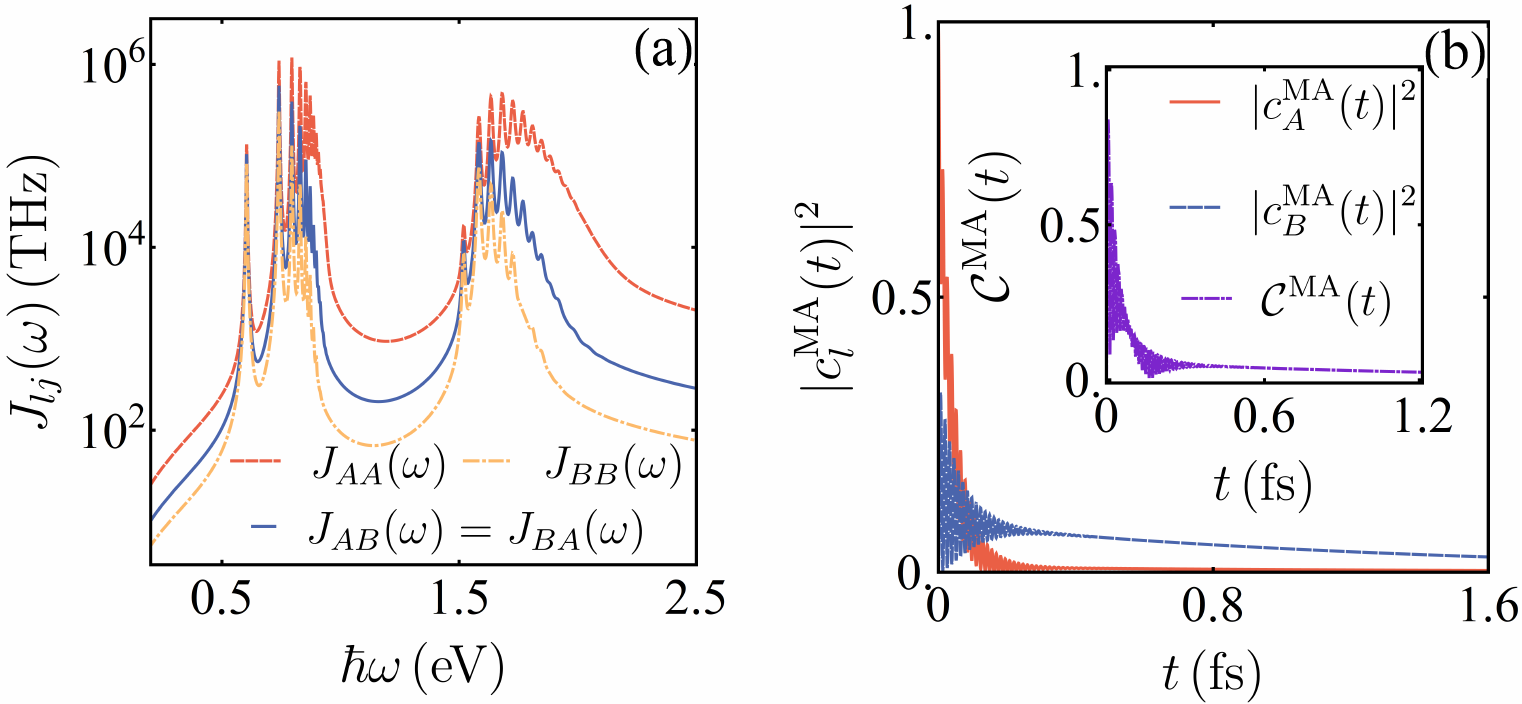}
\caption{(a) Spectral densities $J_{lj}(\omega)$ as a function of the energy. (b) Evolution of the excited-state probability $|c^\text{MA}_l(t)|^2$ under the Markovian approximation. The inset shows the entanglement $\mathcal{C}^\text{MA}(t)$ between the QEs measured by concurrence. We use $\hbar\omega_0=0.7$ eV, $\Gamma_0=20$ THz, $z_A=7$ nm, $z_B=15$ nm, and $R=60$ nm.    }\label{dynk}
\end{figure}

\section{Numerical results}\label{sec:withdissipation}
Via calculating the Green's tensor \eqref{spec} for two QEs located at $z_A=7$ nm and $z_B=15$ nm from the MoS$_2$ nanodisk with radius $R = 60$ nm, we show in Fig. \ref{dynk}(a) the spectral densities $J_{lj}(\omega)$. The sharp peaks representing the resonances between the QEs and EP can be observed. The resonant frequencies are determined by the exciton energies and the structure of the MoS$_2$ nanodisk \cite{PhysRevB.93.035426}. The spectral densities $J_{ll}(\omega)$ relate to the spontaneous emission rate of the QEs renormalized by the MoS$_2$ nanodisk as $\gamma_{ll}=\pi J_{ll}(\omega_0)$ under the Markovian approximation. The enhancement of the spontaneous emission rate by engineering the spatial confinement of the radiation field was conventionally called the Purcell effect \cite{PhysRevLett.112.253601,Toermae2014,PhysRevB.93.035426,Baranov2018,PhysRevB.100.245403,PhysRevB.102.075446,PhysRevResearch.2.033141,PhysRevB.103.035416}. We see that a Purcell factor, i.e., $\gamma_{ll}/\Gamma_0$, about $10^5$ is obtained when the QE-MoS$_2$ distance is $7$ nm. It indicates a dramatically enhanced coupling between the QEs and the EP of the MoS$_2$. The spontaneous emission rate decreases exponentially with the increase of the QE-MoS$_2$ distance, which is consistent with Ref. \cite{ZAYATS2005131}. Choosing the QE frequency as the first resonance peaks, we plot in Fig. \ref{dynk}(b) the Markovian approximate evolution of $|c^\text{MA}_l(t)|^2$ in Eqs. \eqref{cat} and \eqref{cbt}. As a result of the near-field enhancement effect, a significant oscillation of $|c^\text{MA}_l(t)|^2$ manifesting the reversible energy exchange and the strong coupling between the QEs mediated by the EP appears. However, due to the severe absorption of the MoS$_2$ to the EP, $|c^\text{MA}_l(t)|^2$ asymptotically decay to zero. Consequently, the entanglement of the two QEs is completely destroyed in the long-time limit.

\begin{figure}[tbp]
\includegraphics[width=1.0\columnwidth]{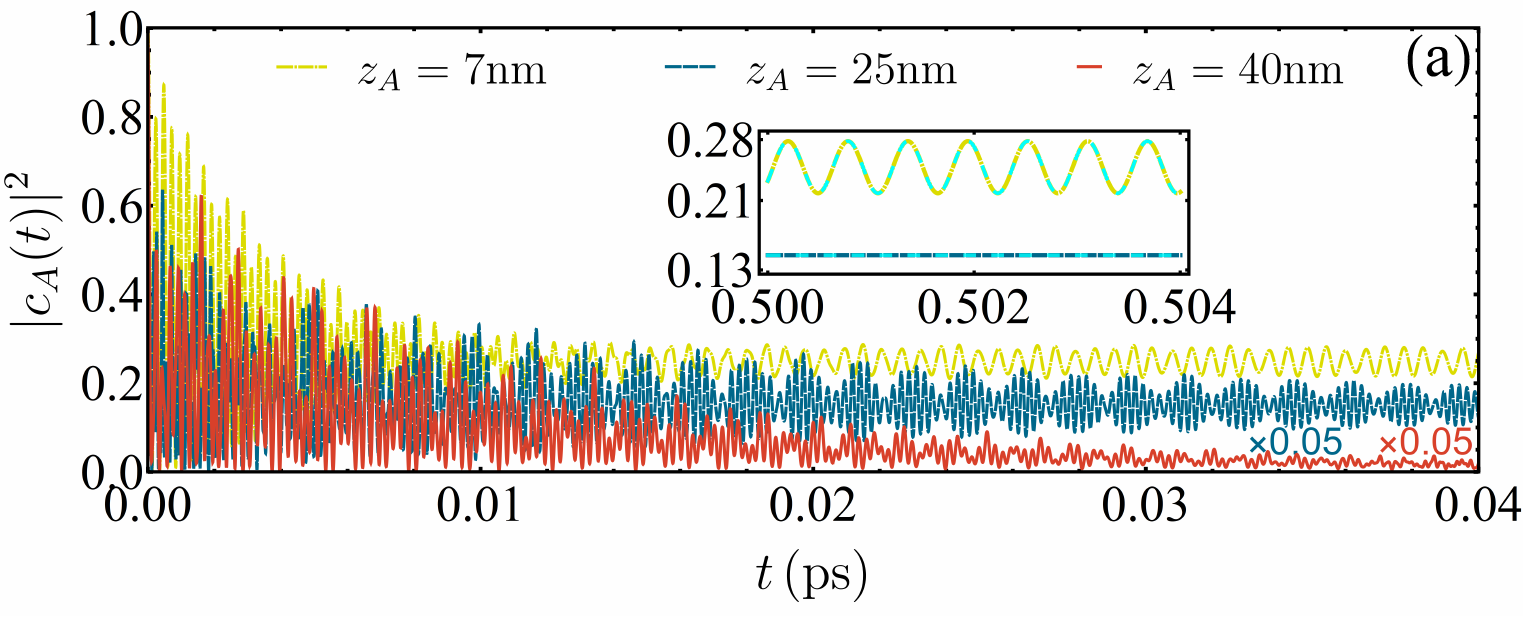}\\
\vspace{0.cm}
\includegraphics[width=\columnwidth]{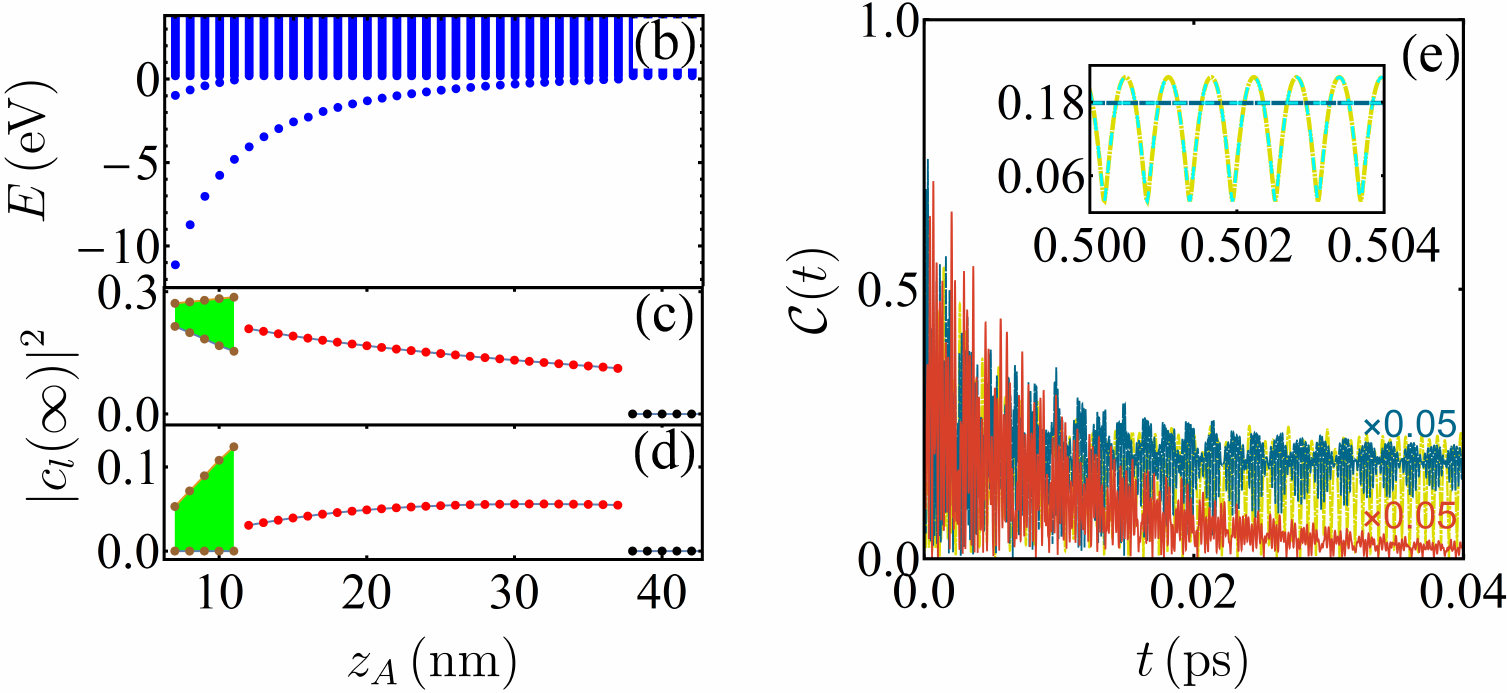}
\caption{ (a) Evolution of $|c_A(t)|^2$ in different $z_A$ obtained by numerically solving Eqs. \eqref{evolution}. (b) Energy spectrum of the whole system in different $z_A$. Two branches of the formed bound states separate the energy spectrum into three regimes with zero, one, and two bound states. Long-time values of $|c_A(\infty)|^2$ (c) and $|c_B(\infty)|^2$ (d) obtained from the exact dynamics (red dots) and from Eq. \eqref{ltmcl} (solid lines). The green region covers the values of $|c_l(\infty)|^2$ during its persistent oscillation. (e) Evolution of concurrence $\mathcal{C}(t)$ obtained by solving Eqs. \eqref{evolution}. The insets of (a) and (e) show the long-time behaviors, where the cyan short dashed lines are obtained from Eq. \eqref{ltmcl}. The evolution time of the blue dashed and red solid lines in (a) and (e) is magnified by a factor $0.05$. Except for $z_B=z_A+8$ nm, the others parameter are the same as those in Fig. \ref{dynk}. }\label{dyn}
\end{figure}

The near-field enhancement of coupling between QEs and the EP makes the weak-coupling requirement of the Markovian approximation break down. To reveal the non-Markovian effect, we numerically solve Eqs. \eqref{evolution} using the spectral densities in Fig. \ref{dynk}(a). Figure \ref{dyn}(a) shows the evolution of $|c_A(t)|^2$ in different $z_A$. The transient dynamics exhibits rapid oscillation due to the energy exchange between the QEs and the EP. We see three kinds of steady-state behaviors in different $z_A$. When the QEs are sufficiently far from the nanodisk, e.g. $z_A=40$ nm, $|c_A(t)|^2$ decays to zero in the long-time limit, which is consistent with the Markovian approximate result in Fig. \ref{dynk}(b). On the contrary, a remarkable difference presents with decreasing $z_A$. It is found that $|c_A(t)|^2$ tends to a nonzero value when $z_A=25$ nm, which denotes a population trapping, while $|c_A(t)|^2$ tends to a lossless oscillation with a constant frequency when $z_A=7$ nm, which is quite like the Rabi oscillation and denotes a persistent energy exchange between the two QEs mediated by the EP. The steady-state values of $|c_A(t)|^2$ match well with the result obtained via Eq. \eqref{ltmcl}; see the insets of Fig. \ref{dyn}(a). It means that the diverse dynamical behaviors in different $z_A$ are essentially determined by the formation of different numbers of bound states. This is further confirmed by the energy spectrum in Fig. \ref{dyn}(b). The two branches of bound-state energies residing in the band-gap regime divide the full spectrum into three regions: without a bound state when $z_A\geq38$ nm, one bound state when $11~\text{nm}< z_A <38$ nm, and two bound states when $z_A\leq11$ nm. The regions perfectly match those where $|c_l(\infty)|^2$ shows different behaviors [see Figs. \ref{dyn}(c) and \ref{dyn}(d)], i.e., complete decay, population trapping, and persistent oscillation, as analytically described by Eq. \eqref{ltmcl}. The bound states are constructive to generate a stable entanglement between the QEs. Different from the asymptotic vanishing under the Markovian approximation in Fig. \ref{dynk}(b) and in the absence of the bound state, the entanglement between the QEs is stabilized in the long-time limit as long as the bound states are formed [see Fig. \ref{dyn}(e)]. Therefore, the dissipation effect of the QEs caused by the absorption of the MoS$_2$ to the EP can be efficiently avoided by engineering the formation of the bound states of the total system. Conventionally, people generally believe that the mediated entanglement of two separated QEs by the polariton modes cannot be preserved to their steady state. To preserve the entanglement, certain extra control ways, for example, a local laser driving on each QE \cite{PhysRevLett.106.020501}, is needed. Our result renews this general belief and proves that, accompanying the formation of the bound states, a stable entanglement can be spontaneously generated by the mediation role of the EP even in the lossy media.

\section{Discussion and conclusions}\label{sec:Conclusion}
Our finding is realizable in the state-of-the-art technique of experiments. The parameters used are near those of the $J$ aggregates as the QEs. Their strong coupling has been studied \cite{PhysRevLett.97.266808,PhysRevLett.108.066401}. It is noted that, although only the specific parameter values of the QEs and MoS$_2$ are considered, our analytically exact result is universal and generalizable to other QEs and semiconductor or even metal nanostructures. Some quantitative differences might occur, but the substantial role played by the bound states in overcoming the dissipation effect of the QEs in the absorbing nanostructure does not change. As a ubiquitous phenomenon, the bound state and its role in the non-Markovian dynamics have been observed in circuit QED \cite{Liu2016} and ultracold atom \cite{Kri2018,Kwon2022} systems. This means that our finding is completely realizable in a quantum EP system, where the strong light-matter coupling is more manifest than in other systems.

In summary, we have investigated the near-field interactions between two QEs and the EP in the monolayer MoS$_{2}$. The sub-diffraction character of the EP endows our system with a strong-coupling nature. A mechanism to overcome the dissipation of the QEs caused by the EP in such an absorbing medium has been revealed. It has been found that as long as one and two bound states are formed for the total system of the QEs and the EP, the dynamics of the QEs exhibit population trapping and persistent Rabi-like oscillation, respectively, with the dissipation efficiently suppressed. Breaking the dissipation barrier of the QEs induced by the EP of the MoS$_2$, our finding supplies an insightful understanding of the light-matter interactions in an absorbing medium.

\section*{Acknowledgments}
This work is supported by the National Natural Science Foundation (Grants No. 11875150, No. 11834005, and No. 12047501) and the Supercomputing Center of Lanzhou University.

\appendix

\section{Electrostatic Green's tensor}\label{appgf}

The Green's tensor induced by an electric dipole in a MoS$_2$ nanodisk is given by ${\bf G}({\bf r},{\bf r}',\omega)={\varepsilon_0c^2\over \omega d}{\pmb \nabla}\phi({\bf r},{\bf r'})$ \cite{PhysRevB.93.035426}, where $\varepsilon_0$ is the vacuum permittivity, $c$ is the speed of light, and $d$ is the dipole moment. The electrostatic potential $\phi({\bf r},{\bf r'})$ is
\begin{eqnarray}
\phi({\bf r},{\bf r'})&=&{R\over 2\varepsilon_0}\sum_{l,n}^\infty \alpha_n^l({\bf r}',\omega)\cos(l\theta)\nonumber\\
&&\times\int_0^\infty{e^{-|z|\over Rp}\mathcal{J}_l(pr)\mathcal{J}_{l+2n+1}(p)\over p}dp,\label{smstvar}
\end{eqnarray}where $R$ is the radius of MoS$_2$, and $\mathcal{J}_k(x)$ is the $k$-order Bessel function of the first kind. The coefficients $\alpha_n^l({\bf r}',\omega)$ forming a vector ${\pmb\alpha}^l=(\alpha_1^l({\bf r}',\omega),\alpha_2^l({\bf r}',\omega),\cdots)^T$ satisfy
\begin{equation}
[-\omega^2{\bf M}^l+\Omega^2(\omega){\bf K}^l]{\pmb \alpha}^l={i\omega\sigma(\omega)\over R^2}{\bf D}^l{\bf d}^l,\label{smmatrx}
\end{equation}where $\Omega^2(\omega)=-i\omega\sigma(\omega)/{2\varepsilon_0 R}$, and the elements of the matrices ${\bf M}^l$, ${\bf K}^l$, and ${\bf D}^l$ are
\begin{eqnarray}
D_{ij}^l&=&{\delta_{ij}\over 2(l+2j+1)},\\
M_{ij}^l&=&{\delta_{i0}\delta_{j0}\over 8l(l+1)^2}+{\delta_{ij}\over4(l+2j)(l+2j+1)(l+2j+2)}\nonumber\\
&&+{\delta_{i+1,j}+\delta_{i,j+1}\over8(l+2j+1)(l+2j+2)(l+2j+3)},\\
K_{ij}^l&=&{\pi^{-1}(-1)^{i-j+1}\over[4(i-j)-1](l+i+j+{1\over 2})(l+i+j+{3\over2})}.
\end{eqnarray}
We consider that the dipole moment of the two QEs at ${\bf r}_l=(0,0,z_l)$ are along the $x$ direction. Thus, only $G_{xx}(z,z',\omega)$ takes effect and only the term $l=1$ contributes to Eq. \eqref{smmatrx}. The elements of ${\bf d}^1$ read
\begin{equation}
d_i^1={(2i+2)d\over 4\pi\varepsilon_0R^2}\int_0^1{r^3\mathcal{P}_i^{(1,0)}(1-2r^2)\over[r^2+({z'\over R})^2]^{3\over 2}}dr,
\end{equation}
where $\mathcal{P}_n^{(a,b)}(x)$ is the Jacobi polynomial. The optical conductivity $\sigma(\omega)$ of MoS$_2$ describes the light-exciton interaction and reads  \cite{PhysRevB.97.205409,PhysRevB.97.205436}
\begin{eqnarray}
\sigma(\omega)&=&\frac{4cv^{2}}{137\pi a_{ex}^{2}\omega}\sum_{k=1,2}\frac{-i}{\omega_{k}-\omega-i\gamma_{k}}\nonumber\\
&&+{m \sigma_{0}\Theta(\omega-\omega_{2})\over\chi}[1+\frac{1+2\beta}{\omega^2/\omega_2^{2}}(1+\beta-\chi)],~~
\end{eqnarray}
where $a_{ex}=0.8$ nm and $\hbar\omega_{2}=3\hbar\omega_1=1.5$ eV are the exciton radius and energies \cite{Basov2016,Chaves2020,PhysRevB.101.161410}, $v=0.55$ nm/fs is the hopping velocity, $\gamma_{1}=2.5$ meV and $\gamma_{2}=5.6$ meV are the damping parameters, $\chi=\sqrt{1+2\beta+\omega^2/\omega_2^{2}}$ with $\beta=0.84$ being a mixing parameter, $\Theta(x)$ is the Heaviside step function, $\sigma_0={e^2\over 16\hbar}$ is the universal optical conductivity, and $m$ is for absorption scaling and set to be one \cite{PhysRevB.91.115407}. With these results, $\alpha_n^1$ is obtained by solving Eq. \eqref{smmatrx}. Then we obtain the Green's tensor at ${\bf r}=(0,0,z)$ by the $x$-direction dipole at ${\bf r}'=(0,0,z')$ as
\begin{equation}
G_{xx}(z,z',\omega)=\frac{-c^{2}}{2\omega^{2}}\sum_{n=0}^{\infty}\alpha_{n}^{1}(z^{\prime},
\omega)\frac{[g(z)-{z\over R}]^{2n+2}}{g(z)},\label{smspec}
\end{equation}
where $g(z)=\sqrt{(z/R)^{2}+1}$ and $\Gamma_{0}=\omega_{0}^{3}d^{2}/(3\pi\hbar\varepsilon_{0}c^{3})$ is the spontaneous emission rate in the free space characterizing the intrinsic lifetime of the dipole.

\section{Markovian approximation}\label{appmark}
Defining $c_{l}(t) =c_{l}^{\prime }(t)e^{-i\omega _{0}t}$, Eqs. \eqref{evolution} convert into
\begin{equation}
\dot{c}'_{l}(t)+\sum_{j=A,B}\int_0^\infty d\omega J_{lj}(\omega )\int_{0}^{t}d\tau e^{-i(\omega-\omega_0)(t-\tau))}c'_{j}(\tau )=0.\label{smevopr}
\end{equation} We make the Markovian approximation $\tilde{c}'(\tau )\simeq \tilde{c}'(t )$ to neglect the memory effect and to extend the upper limit of the $\tau$ integration in Eq. \eqref{smevopr} from $t$ to infinity. It is valid when the QE-EP coupling is weak and the correlation time of the EP is much smaller than the typical time scale of the QEs. Using the equality $\lim_{t\rightarrow\infty}\int_0^t d\tau e^{-ix(t-\tau)}=\pi\delta(x)-i\mathcal{P}\Big({1\over x}\Big)$,
where $\mathcal{P}$ is the Cauchy principal value, we convert Eqs. \eqref{smevopr} into two coupled ordinary differential equations. Their solutions are
\begin{eqnarray}
c^\text{MA}_{A}(t) &=&e^{-[i\omega _{0}+(\Upsilon _{AA}+\Upsilon _{BB})]t}[\frac{\Upsilon _{BB}-\Upsilon _{AA}}{\Omega }\sinh (\Omega t)\nonumber \\
&&+\cosh(\Omega t)],\\
c^\text{MA}_{B}(t) &=&-2e^{-[i\omega _{0}+(\Upsilon _{AA}+\Upsilon _{BB})]t}\frac{%
\Upsilon _{BA}}{\Omega }\sinh (\Omega t),
\end{eqnarray}
where $\Omega =\sqrt{(\Upsilon _{AA}-\Upsilon _{BB})^{2}+4\Upsilon _{AB}\Upsilon_{BA}}$, $\Upsilon _{lj} =\frac{\gamma _{lj}}{2}+i\frac{\Delta _{lj}}{2}$, $\gamma _{lj}=\pi J_{lj}(\omega _{0})$, and $\Delta_{lj}=\mathcal{P}\int d\omega {\frac{J_{lj}(\omega )}{\omega _{0}-\omega }}$.

\section{Energy spectrum}\label{appenspe}
The eigenstate for the total QE-EP system is expanded as $|\Phi\rangle=\big[\sum_{l}x_l\hat{\sigma}_l^\dag+\int d^{3}\mathbf{r}\int d\omega y_{{\bf r},\omega}\hat{\mathbf{f}}^\dag(\mathbf{r},\omega)\big]|g_A,g_B;\{0_{{\bf r},\omega}\}\rangle$. Then, we derive from $\hat{H}|\Psi\rangle=E|\Phi\rangle$ that
\begin{eqnarray}
(E-\hbar\omega_0)x_{l} &=&-\int d^{3}\mathbf{s}\int d\omega ^{\prime
}y_{\mathbf{s},\omega ^{\prime }}\frac{ic^{-2}\omega ^{\prime 2}}{\sqrt{\pi
\varepsilon _{0}/\hbar }}\nonumber\\
&&\times\sqrt{\text{Im}[\varepsilon (\omega ^{\prime })]}\mathbf{d}%
_{l}\cdot \mathbf{G}(\mathbf{r}_{l},\mathbf{s},\omega ^{\prime }),\label{smaba}\\
(E-\hbar\omega')y_{\mathbf{s},\omega ^{\prime }} &=&\frac{i\omega ^{\prime 2}\sqrt{\text{Im}[\varepsilon
(\omega ^{\prime })]}}{c^{2}\sqrt{\pi \varepsilon _{0}/\hbar }}\nonumber\\
&&\times\sum_{m=A,B}x_{m}\mathbf{d}%
_{m}\cdot \mathbf{G}^{\ast }(\mathbf{r}_{m},\mathbf{s},\omega ).\label{smbss}
\end{eqnarray}
Substitution $y_{\mathbf{s},\omega ^{\prime }}$ from Eq. \eqref{smbss} into Eqs. \eqref{smaba} and using $\int d^3\mathbf{s}\frac{\omega ^{2}}{c^{2}}$Im$[\varepsilon(\omega )]\mathbf{G}(\mathbf{r},\mathbf{s},\omega )\mathbf{G}^{\ast }(\mathbf{r}^{\prime },\mathbf{s},\omega )=\textrm{Im}[\mathbf{G}(\mathbf{r},\mathbf{r}^{\prime },\omega )]$ result in
\begin{eqnarray}
(E-\hbar\omega_0)x_{l}&=&\hbar ^{2}\int d\omega \frac{\sum_{m=A,B}J_{lm}(\omega
)x_{m}}{E-\hbar \omega },~\label{smxba}
\end{eqnarray}
where $J_{lm}(\omega )=\frac{\omega ^{2}}{\pi \hbar \varepsilon _{0}c^{2}}%
\mathbf{d}_{l}\cdot \text{Im}[\mathbf{G}(\mathbf{r}_{l},\mathbf{r}%
_{m},\omega ^{\prime })]\cdot \mathbf{d}_{m}$.
Eliminating $x_A$ and $x_B$ in Eqs. \eqref{smxba}, we have
\begin{equation}
\Xi _{A}({E\over i\hbar})\Xi _{B}({E\over i\hbar})-\tilde{f}_{AB}({E\over i\hbar})\tilde{f}_{BA}({E\over i\hbar})=0,\label{aadd}
\end{equation}
where $\Xi _{l}(s)=s+i\omega _{0}+\tilde{f}_{ll}(s)$ and $\tilde{f}_{lj}(s)=\int_{0}^{\infty } {J_{lj}(\omega)d\omega\over s+i\omega}$.
The solution of Eq. \eqref{aadd} gives the eigenenergies of the total system.

\section{Steady-state solutions}\label{appsssn}
The time-dependent $c_l(t)$ is solvable via the inverse Laplace transform to $\tilde{c}_l(s)$ in Eq. \eqref{cldtts}, i.e.,
\begin{equation}
c_l(t)={1\over 2\pi i}\int_{\sigma-i\infty}^{\sigma+i\infty}\tilde{c}_l(s)e^{st}ds,\label{smcltd}
\end{equation}which can be done by finding the poles from
\begin{equation}
\Xi _{A}(s)\Xi _{B}(s)-\tilde{f}_{AB}(s)\tilde{f}_{BA}(s)=0.\label{emeigroot}
\end{equation}
$\tilde{f}_{lj}(s)$ is divergent for any negative pure-imaginary $s$, which forms a branch cut for Eq. \eqref{smcltd}. Due to the quadratic-like nature of Eq. \eqref{emeigroot} for positive pure-imaginary $s$, $\tilde{c}_l(t)$ has two isolated poles at most in the positive imaginary axis of $s$.

\begin{figure}[tbp]
\includegraphics[width=.45\columnwidth]{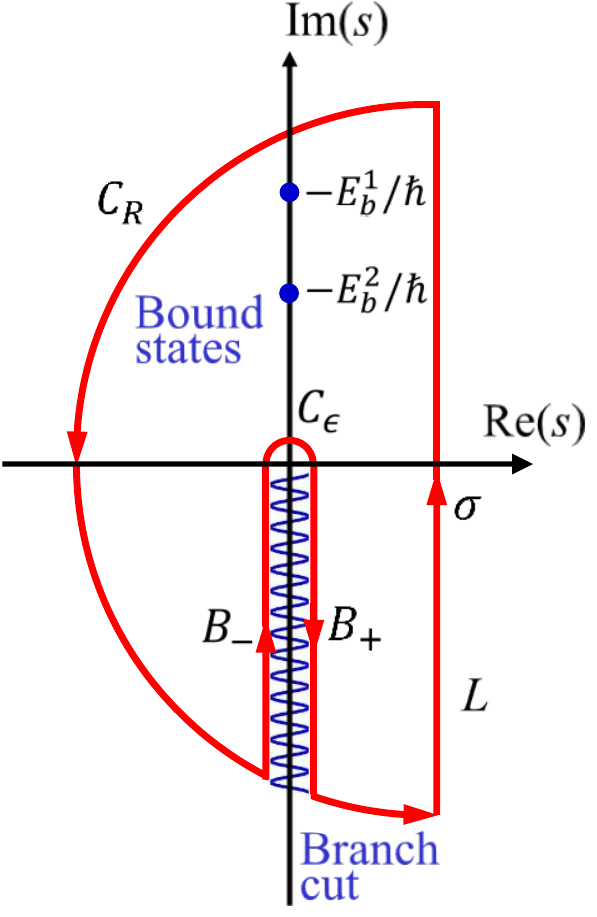}
\caption{Path of the contour integration in the complex plane of $s$ to calculate the inverse Laplace transform of $\tilde{c}_l(s)$. Two poles at most are present in the positive imaginary axis of $s$.}\label{ctfg}
\end{figure}

Figure \ref{ctfg} shows the path of the contour integration to evaluate Eq. \eqref{smcltd}. According to the residue theorem, we have
\begin{eqnarray}
&&\lim_{\epsilon\rightarrow 0,R\rightarrow\infty}\left[\int_{\sigma-iR}^{\sigma+iR}+\int_{C_R}+\int_{B_-}+\int_{C_\epsilon}+\int_{B_+}\right]\tilde{c}_l(s)e^{st}ds\nonumber\\
&&=2\pi i\sum_j\text{Res}({E_b^j\over i\hbar}),
\end{eqnarray}
where $\text{Res}({E_b^j\over i\hbar})$ is the residue contributed by the $j$th bound state. From Jordan's lemma, the integration along $C_R$ and $C_\epsilon$ is
negligible. The paths making a contribution to $c_l(t)$ are $B_\pm$, i.e.,
\begin{eqnarray}
c_l(t)=\sum_j\text{Res}_l({E_b^j\over i\hbar})-\lim_{\epsilon\rightarrow 0}\left[\int_{B_-}+\int_{B_+}\right]{\tilde{c}_l(s)e^{st}\over 2\pi i}ds.~~~~~~\label{smsclt}
\end{eqnarray}
The integral in Eq. \eqref{smsclt} tends to zero in the long-time limit due to the out-of-phase interference of the integral variable $s$ along the negative imaginary axis. Therefore, we do not bother to calculate them if only the steady-state form of $c_l(t)$ is concerned. The $j$th residue of $c_l(t)$ can be evaluated by the L'Hospital's rule as $\text{Res}_l({E_b^j\over i\hbar})=Z_l^je^{{-i\over\hbar}E_b^jt}$, where $Z_l^j=\frac{\zeta _l(s)}{\partial_s[\Xi _{A}(s)\Xi _{B}(s)-\tilde{f}_{AB}(s)\tilde{f}_{BA}(s)]}\big|_{s=E_{b}^{j}/(i\hbar) }$.
We finally obtain the steady-state form of $c_l(t)$ as Eq. \eqref{ltmcl}.

\bibliography{Ref}
\end{document}